\documentstyle[12pt]{article}

\begin{document}

\begin{flushright}
IMSc/2009/09/13
\end{flushright} 

\vspace{2mm}

\vspace{2ex}

\begin{center}

{\large \bf Particle Motion with Ho\v{r}ava -- Lifshitz type} \\

\vspace{2ex}

{\large \bf Dispersion Relations}

\vspace{8ex}

{\large S. Kalyana Rama}

\vspace{3ex}

Institute of Mathematical Sciences, C. I. T. Campus, 

Tharamani, CHENNAI 600 113, India. 

\vspace{1ex}

email: krama@imsc.res.in \\ 

\end{center}

\vspace{6ex}

\centerline{ABSTRACT}
\begin{quote} 

Using super Hamiltonian formalism, we study the motion of
particles whose dispersion relations are modified to incorporate
Ho\v{r}ava -- Lifshitz type anisotropic scaling symmetry. We
find the following as consequences of this modified dispersion
relation: (i) The speed of a charged particle under a constant
electric field grows without bound and diverges. (ii) The speed,
as measured by an observer locally at rest, of a particle
falling towards the horizon also grows without bound and
diverges as the particle approaches the horizon. (iii) This
particle reaches the horizon in a finite coordinate time, in
contrast to the standard case where it requires infinite time.

\end{quote}

\vspace{2ex}







\newpage

\vspace{2ex}

\centerline{ {\bf 1. Introduction} }

\vspace{2ex}

Recently, Ho\v{r}ava has proposed \cite{h1} a candidate theory
for gravity, symmetric under Lifshitz type anisotropic scaling
of spacetime coordinates
\begin{equation}\label{z}
x^i \; \to \; l x^i
\; \; \; , \; \; \; 
t \; \to \; l^z t 
\end{equation}
where $z$ is the scaling exponent. This theory is renormalisable
if $z = d$ and superrenormalisable if $z > d$ where $d$ is the
number of spatial dimensions, and has many appealing features
which are being actively explored. See \cite{ cosmology,
blackholes, motion, misc} for a sample of references.

Some of the features of Ho\v{r}ava's theory are: it generically
leads to a dispersion relation of the form
\begin{equation}\label{dr0} 
\omega^2 - \left( c^2 k^2 + \cdots + \alpha (k^2)^z \right) = 0
\end{equation} 
for massless excitations where $c$ is the speed of light in
vacuum in the IR and $\alpha$, assumed to be $> 0 \;$, is a
marginal coupling parameter. The speed of propagation diverges
in the UV. In early universe, such a dispersion relation leads
to scale invariant spectrum of fluctuations, and also to matter
equation of state $p = \frac{z} {d} \; \rho \;$. Ho\v{r}ava's
theory may also lead to a bounce in the scale factor evolution
if the spatial curvature is non zero. These and a variety of
other cosmological consequences have been explored in
\cite{cosmology}.

Static spherically symmetric solutions of Ho\v{r}ava's theory
have also been studied. Such solutions generically have
horizon(s) but differ from the standard Schwarzschild solution
in other details. The thermodynamical properties of such
solutions have also been studied. See \cite{blackholes} for a
sample of references. Geodesic motion of probes in such
spacetimes have also been studied \cite{motion}, as also a host
of other issues \cite{misc}.

Given the generic modifications of dispersion relations, it is
likely that those for classical particles, or semiclassical
probe wave packets, in Ho\v{r}ava's theory are also modified to
a form of the type
\begin{equation}\label{dr1}
E^2 - \left( c^2 p^2 + \cdots + \alpha (p^2)^z \right) 
= m^2 c^4 
\end{equation}
which incorporates Ho\v{r}ava -- Lifshitz type anisotropic
scaling symmetry (\ref{z}) where $z = d$ for renormalisable case
and $z > d$ for superrenormalisable case. The particle dynamics
will then differ from the standard one for high momentum $p \;$,
equivalently high energy $E \;$. Also, in the presence of
gravitational field, probes obeying such dispersion relations
are unlikely to follow the spacetime geodesics. It is clearly of
interest to study the dynamics of such particles and the
consequent differences from the standard case.

In this paper, we study the motion of particles whose dispersion
relation is modified to incorporate Ho\v{r}ava -- Lifshitz type
anisotropic scaling symmetry (\ref{z}). It turns out that the
required modifications suggest themselves naturally not in the
Lagrangian formalism, \footnote{However, see \cite{romero} where
anisotropic scaling transformations are implemented in the
context of `conformal mechanics' using Lagrangian formalism.}
but in the super Hamiltonian formalism described in \cite{mtw}.
Equations of motion, with or without gravitational fields, may
then be easily obtained. \footnote{ Other types of dispersion
relations, which are typically of the form $ F(E, p^2) =
constant \;$ \cite{transplanck, dispersion}, may also be
analysed similarly; the relevent equations may be obtained
straightforwardly, but will not be presented here.}

Using this formalism, we first obtain the equations of motion
for a charged particle without gravitational fields and study
the consequences when only a constant electric field is
present. This will illustrate the main feature of the dynamics,
which one expects from the relation (\ref{dr1}), that the speed
of particle grows without bound and diverges for high momentum
$p \;$, equivalently high energy $E \;$. 

We then consider gravitational fields, obtain the equations of
motion, specialise to spacetimes with horizon, and analyse the
motion outside the horizon. Here again we find that the speed of
the particle, as measured by an observer locally at rest
\cite{ll, ct}, grows without bound and diverges as the particle
approaches the horizon. Also, we find that the particle reaches
the horizon in a finite coordinate time, in contrast to the
standard case where it requires infinite coordinate time.

The plan of the paper is as follows. In section 2, we present
relevant details of super Hamiltonian formalism, incorporate
anisotropic scaling symmetry, and present the general equations
of motion. In section 3, we study the motion of a charged
particle without gravitational fields. In section 4, we study
the motion with gravitational fields present. In section 5, we
conclude with a few remarks.

\vspace{2ex}

\centerline{ {\bf 2. Formalism} }

\vspace{2ex}

We first set up our notation to describe particle motion obeying
a given modified dispersion relations. We will use the super
Hamiltonian formalism given in \cite{mtw} and write the
equations of motion in a general form, which can then be
analysed for specific cases.

Consider the motion of a particle in $(3 + 1)$ -- dimensional
spacetime. Let $x^\mu = (x^0, x^i) = (t, x^i) \;$, $i = 1, 2, 3
\;$, denote spacetime coordinates, let $\lambda$ parametrise the
particle trajectory, and let $u^\mu = \frac{d x^\mu}{d \lambda}
\;$. Equations of motion can be obtained in a standard way by
starting with a Lagrangian $L(u^\mu, x^\nu) \;$. However,
incorporating modified dispersion relations in the Lagrangian is
not straightforward. Hence we follow an equivalent, alternate
formalism \cite{mtw} in which one starts with a `super
Hamiltonian' ${\cal H}(p_\mu, x^\nu) \;$ where $p_\mu$ is the
momentum, and obtains the equations of motion by extremising the
action
\begin{equation}\label{smtw} 
S = \int \left( \; p_\mu d x^\mu - {\cal H} \; d \lambda \;
\right)
\end{equation}
with respect to $x^\mu$ and $p_\mu \;$. The resulting equations 
of motion are given by 
\begin{equation}\label{eom}
u^\mu \equiv \frac{d x^\mu}{d \lambda} = 
\frac{\partial {\cal H}} {\partial p_\mu} \; \; , \; \; \;
\frac{d p_\mu}{d \lambda} = 
- \frac{\partial {\cal H}} {\partial x^\mu} \; \; .
\end{equation}
These equations imply that $\frac{d {\cal H}} {d \lambda} = 0
\;$, hence that ${\cal H} = {\cal H}_0 = constant \;$, along the
particle trajectory. A straightforward exercise now shows that
the standard geodesic equations can be obtained from
\begin{equation}\label{h0}
{\cal H} = {\cal H}_{std} \equiv \frac{c^2}{2} 
\; G^{\mu \nu} (x) \; p_\mu p_\nu
\end{equation}
where $G^{\mu \nu}$ is the inverse of the spacetime metric
$G_{\mu \nu}$. In the case of charged particles, one first
replaces $p_\mu$ by $(p_\mu + e a_\mu)$ in ${\cal H} \;$ where
$e$ is the particle charge, and $a_\mu$ is the $4-$vector
potential, and then extremises the action $S \;$. The constant
of motion ${\cal H}_0 = 0 \;$ for massless particles and, in our
notation, ${\cal H}_0 < 0 \;$ for particles with non zero rest
mass. See \cite{mtw} for further details. Also, in our notation,
the dimensions of $p_0 \;$, ${\cal H} \;$, and $\lambda \;$ are
$(energy)$, $(energy)^2$, and $(time/energy) \;$ respectively as
follows from $x^0 = t$ and equations (\ref{smtw}) and
(\ref{h0}).

In Ho\v{r}ava's theory, the gravitational fields $(N, N^i, g_{i
j}) \;$ are taken to be those of the metric $G_{\mu \nu}$ given
by
\begin{equation}\label{ds}
d s^2 = G_{\mu \nu} d x^\mu d x^\nu = - c^2 N^2 d t^2 
+ g_{k l} (d x^k + N^k d t) (d x^l + N^l d t) \; \; . 
\end{equation}
The anisotropic scaling dimensions of various quantities in
momentum units are given, taking $x^i$ to have length
dimensions, by
\begin{eqnarray}
& & [x^i] = - 1 \; \; , \; \; \; [t] = - z \; \; , \; \; \;
[p_0] = z \nonumber \\
& & [c] = z - 1 \; \; , \; \; \; [{\cal H}] = - [\lambda] = 2 z
\nonumber \\
& & [N] = [g_{i j}] = 0 \; \; , \; \; \; 
[N^i] = z - 1 \; \; . \label{1}
\end{eqnarray}

Using equation (\ref{ds}), ${\cal H}_{std} \;$ may now be
written as
\begin{equation}\label{hstd} 
{\cal H}_{std} = \frac{1}{2} \; \left( - \;  
\frac{(p_0 - N^k p_k)^2}{N^2} + c^2 \; g^{k l} p_k p_l \right)
\end{equation}
which leads to the flat-space dispersion relation $E^2 - c^2 p^2
= constant \;$. The standard super Hamiltonian ${\cal H}_{std}
\;$ can now be generalised to incorporate any modified
dispersion relation of the form $ F(E, p^2) = constant \;$
\cite{ transplanck, dispersion}. A natural proposal for such a
generalised super Hamiltonian is given by
\begin{equation}\label{hgen}
{\cal H} = \frac{1}{2} \; F(\epsilon, X) \; \; ; \; \; \;
\epsilon = \sqrt{\frac{(p_0 - N^k p_k)^2}{N^2}} \; \; , \; \; \; 
X = g^{k l} p_k p_l \; \; .
\end{equation}
The modified equations of motion for particle trajectory, with
or without gravitational fields, may now be obtained using
equations (\ref{eom}) and (\ref{hgen}). Those for charged
particle may be obtained similarly upon replacing $p_\mu$ by
$(p_\mu + e a_\mu)$ in ${\cal H} \;$.

Although it is straightforward to analyse the general case, we
will consider here only Ho\v{r}ava -- Lifshitz type modified
dispersion relations as in equation (\ref{dr1}). The
corresponding super Hamiltonian is naturally given by
\begin{equation}\label{hhl}
{\cal H} = \frac{1}{2} \; \left( - \; 
\frac{(p_0 - N^k p_k)^2}{N^2} + f(X) \right) \; \; , \; \; \;
f(X) = \sum_{n = 1}^z \alpha_n X^n
\end{equation}
where $X = g^{k l} p_k p_l \;$, $z = 3$ for renormalisable case,
and $z > 3$ for superrenormalisable case. The scaling dimensions
of $\alpha_n \;$ are given by $[\alpha_n] = 2 (z - n) \;$. Thus,
$\alpha_z \;$ has vanishing scaling dimensions and is a marginal
coupling parameter; other $\alpha_n$ have positive scaling
dimensions and are relevant parameters. We set $\alpha_1 = c^2
\;$ so as to obtain the standard case in the IR where $X \;$ is
small. Determining other $\alpha_n \;$ requires detailed
knowledge of the theory which describes the particles, or
semiclassical probe wave packets.  Here, for simplicity, we just
assume that $\alpha_n \ge 0 \;$ and that $\alpha_z \equiv \alpha
> 0 \;$. This will be sufficient for our purposes here.

Before proceeding further, we now make a remark. In principle,
given the super Hamiltonian ${\cal H} \;$, it is possible to
obtain the corresponding Lagrangian $L$ by an appropriate
Legendre transformation. However, the procedure involves
inversion of a function which we are unable to carry out
explicitly in a closed form except in very simple cases. The
explicit form of the resulting $L$ is not illuminating, nor does
it suggest a natural generalisation of $L \;$ which may
incorporate a given modified dispersion relation.

\vspace{2ex}

\centerline{ {\bf 3. Charged particle motion in flat space} }

\vspace{2ex}

For the purpose of illustrating the effects of modified
dispersion relations, we first consider the motion of a charged
particle in flat spacetime. Thus, $(N, N^i, g_{i j}) = (1, 0,
\delta_{i j}) \;$ in equation (\ref{ds}), and $p_\mu$ is to be
replaced by $(p_\mu + e a_\mu)$ in ${\cal H} \;$ where $e$ is
the particle charge and $a_\mu$ is the $4-$vector potential.
The super Hamiltonian in equation (\ref{hhl}) thus becomes
\begin{equation}\label{hhlem}
{\cal H} = \frac{1}{2} \; \left( - (p_0 + e a_0)^2 + f(X)
\right) \; \; , \; \; \; f(X) = \sum_{n = 1}^z \alpha_n X^n
\end{equation}
where $X = \delta^{k l} (p_k + e a_k) (p_l + e a_l)\;$. Let
\begin{eqnarray*}
\eta_{\mu \nu} = diag \;(- c^2, 1, 1, 1) & , & 
F_{\mu \nu} = \partial_\mu a_\nu - \partial_\nu a_\mu \\
\left( \hat{u}^0, \; \hat{u}^i \right) 
= \left(\frac{u^0}{c^2}, \; \frac{u^i}{f'} \right)
& , & f' = \frac{d \; f}{d X} \; \; , 
\end{eqnarray*}
and $\eta^{\mu \nu}$ be the inverse of $\eta_{\mu \nu} \;$.
Then, after a straightforward algebra using equations
(\ref{eom}) and ${\cal H} = {\cal H}_0 \equiv - \frac{1}{2} m^2
c^4 \;$ where $m$ is the rest mass of the particle, it follows
that the charged particle motion is described by
\begin{eqnarray}
\hat{u}^\nu & = & \eta^{\mu \nu} \; (p_\mu + e a_\mu) 
\nonumber \\
X & = & \delta_{k l} \; \hat{u}^k \; \hat{u}^l \nonumber \\
(u^0)^2 & = & m^2 c^4 + f(X) \nonumber \\
\frac{d \; \hat{u}^\mu}{d \lambda} & = & e \; 
\eta^{\mu \alpha} \; u^\beta \; F_{\beta \alpha} \; \; .
\label{emeom}
\end{eqnarray}
The velocity $v^i$ with respect to the observer time $t \;$ is
given by $v^i = \frac{d x^i}{d t} = \frac{u^i}{u^0} \;$. It
then follows from the above expressions that 
\begin{equation}\label{vt}
v^2 = \delta_{k l} \; v^k \; v^l = 
\frac{X \; (f')^2}{m^2 c^4 + f(X)} 
\; \; \to \; \; \alpha_z \; z^2 \; X^{z - 1} 
\end{equation}
in the limit $X \to \infty \;$ for the function $f(X) \;$ in
equation (\ref{hhlem}).

To illustrate the salient features of the modified dispersion
relation, let us consider the case where there is a constant
electric field $E$ along $x^1 -$ direction, {\em i.e.} $F_{0 1}
= E \;$ and other independent components of $F_{\mu \nu} = 0
\;$. Let $u^i = 0 \;$, and hence $X = f(X) = 0 \;$, initially at
$t = \lambda = 0 \;$. It is then clear that $u^2 = u^3 = 0 \;$
for all $t > 0 \;$. Also, let $u^1 = u \;$ and $\hat{u} =
\frac{u}{f'} \;$. Equations (\ref{emeom}) now become
\begin{equation}\label{emeom2}
\frac{d \; \hat{u}}{d \lambda} = e \; E \; u^0 \; \; , \; \; \;
(u^0)^2 = m^2 c^4 + f(X) \; \; , \; \; \; X = \hat{u}^2 \; \; . 
\end{equation}
Now $\lambda (\hat{u}) \;$ and, thereby in principle,
$\hat{u}(\lambda) \;$ follows from
\begin{equation}\label{lambdauem}
e \; E \; \lambda (\hat{u})  
= \int_0^{\hat{u}} \frac{d \; \hat{u}} {u^0} 
= \int_0^{\hat{u}} \frac{d \; w} 
{\sqrt{ m^2 c^4 + f(w^2)}} \; \; .
\end{equation}
Then, $X(\lambda) \;$ and $u^0 (\lambda) \;$ may be obtained
using equations (\ref{emeom2}). 

To proceed further, we need the explicit form of $f(X) \;$. Let
$f(X) = c^2 X \;$. In this case, it can be shown easily that the
standard answers are obtained. That is, $e E t = m c \; Sinh \;
(e c E \lambda) \;$, the characteristic time is $\frac{m c} {e
E} \; $, and
\begin{equation}
u^2 = c^4 X = (e c^2 E t)^2 \; \; , \; \; \; 
(u^0)^2 = m^2 c^4 + (e c E t)^2 \; \; .
\end{equation}
From equation (\ref{vt}), and from the above expressions, it
also follows that the velocity $v = \frac{u}{u^0} \to c \;$ in
the limit $t \to \infty \;$, as expected of a particle acted
upon by a constant force. Also, note that $\tau = (m c^2) \;
\lambda \;$ is the standard proper time of the particle since
equation (\ref{emeom2}) for $(u^0)^2 \;$, written in terms of
$\tau \;$, becomes
\[
\eta_{\mu \nu} \; \frac{d x^\mu}{d \tau} \; 
\frac{d x^\nu}{d \tau} = - c^2 \; \; . 
\]

Now consider $f(X) = \alpha X^z \;$ with $z > 1 \;$. Then
$\lambda(\hat{u}) \; $ in equation (\ref{lambdauem}) approaches
a finite value $\lambda_* \;$ from below as $\hat{u} \to \infty
\;$. Therefore, $\hat{u} (\lambda) \;$ and hence $X(\lambda) \;$
diverge to $\infty \;$ as $\lambda \to \lambda_{* -} \;$.
Equation (\ref{vt}) then shows that, in this limit, the velocity
$v \to \infty \;$ since $z > 1 \;$.

It is easy to obtain asymptotic expressions for various
quantities in the limit $\lambda \to \lambda_{* -} \;$. For
example,
\[
t \sim (\lambda_* - \lambda)^{- \frac{1}{z - 1}} 
\; \; , \; \; \; 
X \sim t^2 \; \; , \; \; \; v^2 \sim t^{2 (z - 1)} \; \; ,
\]
all diverging to $\infty \;$ in the limit $\lambda \to
\lambda_{* -} \;$ since $z > 1 \;$. Thus we see that, as a
consequence of the Ho\v{r}ava -- Lifshitz type modified
dispersion relation, the speed of a particle which is acted upon
by a constant force increases without bound and diverges to
$\infty \;$ in the limit $t \to \infty \;$. There is no
inconsistency here because the particle is in the UV regime in
this limit since $X \;$ is large and, in Ho\v{r}ava's theory,
there is no Lorentz invariance in the UV.

\vspace{2ex}

\centerline{ {\bf 4. Particle motion under gravity} }

\vspace{2ex}

We now consider the motion of a particle in spacetime with non
trivial gravitational fields $(N, N^i, g_{i j}) \;$ as in
equation (\ref{ds}). For convenience, we set $N^2 = B \;$ and
assume that $B > 0 \;$. Equations of motion are obtained using
equations (\ref{eom}) and (\ref{hhl}). With $X = g^{k l} p_k p_l
\;$ and $f' = \frac{d \;f}{d X} \;$, they are given by
\begin{eqnarray} 
u^0 & = & - \; \frac{p_0 - N^k p_k}{B} \label{u0p} \\ 
u^i & = & f' \; g^{i k} p_k 
+ \frac{(p_0 - N^k p_k)}{B} \; N^i \label{uip} \\
{\cal H} & = & \frac{1}{2} \; \left( 
- \frac{(p_0 - N^k p_k)^2}{B} + f(X) \right) = {\cal H}_0
\label{hh0}
\end{eqnarray} 
and
\begin{eqnarray} 
\frac{d \; p_\mu}{d \lambda} & = & - \; 
\frac{(p_0 - N^k p_k)^2}{2 \; B^2} \; \partial_\mu B 
- \frac{1}{2} \; f' \; p_k p_l \; \partial_\mu g^{k l} 
\nonumber \\
& & - \; \frac{(p_0 - N^k p_k)}{B} \; p_l \; \partial_\mu N^l 
\; \; . \label{pmulambda}
\end{eqnarray} 
One can now express $p_\mu \;$, thereby $X$ and ${\cal H} \;$,
in terms of $u^\mu \;$ using above equations. Thus,
\begin{eqnarray} 
p_i & = & \frac{1}{f'} \; g_{i k} \; (u^k + N^k u^0) 
\label{piu} \\
p_0 & = & - B u^0 + \frac{1}{f'} \; g_{k l} \; N^k \; 
(u^l + N^l u^0) \label{p0u} \\
X & = & \frac{1}{f'^2} \; g_{k l} \; (u^k + N^k u^0) \; 
(u^l + N^l u^0) \label{xu} \\
{\cal H} & = & \frac{1}{2} \; \left( 
- B \; (u^0)^2 + f(X) \right) = {\cal H}_0 \; \; . \label{hu}
\end{eqnarray}
It can be shown after some algebra that, for $f(X) = c^2 X \;$,
the above equations of motion reduce to the standard geodesic
equations corresponding to the metric given in equation
(\ref{ds}).

The initial conditions may be taken to be the values of $x^\mu
\;$ and $u^\mu \;$ initially at $t = \lambda = 0 \;$. Then, in
principle, the initial values of $X \;$ and $p_\mu \;$, and the
constant of motion ${\cal H}_0 \;$ may all be obtained from
equations (\ref{p0u}) -- (\ref{hu}). Further evolution is then
determined by the equations of motion.

A general feature of the particle motion may now be seen
directly from the above equations. First, using equations
(\ref{ds}), (\ref{xu}) and (\ref{hu}), we write
\begin{equation}\label{umuunu}
G_{\mu \nu} u^\mu u^\nu = 
2 \; c^2 \; {\cal H}_0 - c^2 \; f(X) + X \; f'^2 \; \; . 
\end{equation}
If $f(X) = c^2 X$ then $G_{\mu \nu} u^\mu u^\nu = 2 c^2 {\cal
H}_0 = constant \;$ along the particle trajectory. If $f(X) \;$
is as given in equation (\ref{hhl}) then, for $z > 1 \;$ and in
the limit $X \to \infty \;$, we have
\[
G_{\mu \nu} u^\mu u^\nu \; \sim \; \alpha X^z \; 
(\alpha z^2 X^{z - 1} - c^2) \; \to \; \infty  \; \; . 
\]

Now, $G_{\mu \nu} u^\mu u^\nu = 0 \;$ for null trajectories and,
in our notation, is negative (positive) for timelike (spacelike)
trajectories. Let the initial conditions be chosen to correspond
to a timelike trajectory {\em i.e.} $G_{\mu \nu} u^\mu u^\nu < 0
\;$ initially at $t = \lambda = 0 \;$. If $X \to \infty \;$ as
the motion evolves then it follows that, for the dispersion
relation in equation (\ref{hhl}) with $z > 1 \;$, $G_{\mu \nu}
u^\mu u^\nu \;$ switches sign, becomes positive and diverges to
$\infty \;$.  This implies that as the motion evolves into the
UV, the initially timelike particle trajectory becomes null and
then spacelike.

This indeed happens, for example, in the spacetime described by
equation (\ref{ds1}) below where $N^i = 0 \;$, the fields are
time independent, $B(r) > 0 \;$ for $r > r_h \;$, and $B(r) \to
0 \;$ as $r \to r_h \;$. Because of time independence, it
follows that $p_0 = constant \;$. Equation (\ref{hh0}) then
gives
\[
f(X) = 2 \; {\cal H}_0 + \frac{p_0^2}{B} 
\]
from which it follows that $f(X) \;$ and, hence, $X \;$ diverge
to $\infty \;$ as an infalling particle approaches $r_h \;$. If
such a particle obeys the modified dispersion relation given in
equation (\ref{hhl}) then we have $z > 1 \;$, and that the
initially timelike particle trajectory becomes null and then
spacelike as the particle approaches $r_h \;$.

\vspace{1ex}

\centerline{ {\bf Static spherically symmetric black hole} }

\vspace{1ex}

We now study the particle motion in the region outside the
(outer) horizon $r_h \;$ of a static spherically symmetric black
hole. We take the line element $d s \;$ to be given by
\begin{equation}\label{ds1}
d s^2 = - c^2 B d t^2 + \frac{d r^2}{B} + {\cal C}
\left( d \theta^2 + sin^2 \theta \; d \phi^2 \right) 
\end{equation}
where $B \;$ and ${\cal C} \;$ are functions of $r \;$ only,
$B(r_h) = 0 \;$, and $B > 0 \;$ for $r > r_h \;$. We study the
motion outside the horizon, {\em i.e.} for $r > r_h \;$.

Since the fields, and hence the super Hamiltonian ${\cal H} \;$,
are independent of $t$ and $\phi$ coordinates, it follows that
$p_0 \;$ and $p_\phi \;$ are constants. Furthermore, let $\theta
= \frac{\pi}{2} \;$ and $u^\theta = 0 \;$ initially. Then we
obtain $p_\theta = 0 \;$. The remaining equations may be written
as
\begin{eqnarray}
B \; u^0 & = & - \; p_0 \; \; , \; \; \; 
{\cal C} \; u^\phi = f' J \label{u0} \\
(u^r)^2 & = & B \; f'^2 \; 
( X - \frac{J^2}{{\cal C}} ) \label{ur} \\
f(X) & = & 2 \; {\cal H}_0 + \frac{p_0^2}{B} \label{fh0} 
\end{eqnarray}
where $p_0 \;$, $J \;$, and ${\cal H}_0 \;$ are constants of
motion. 

The parameter $J$ describes orbital motions. The allowed range
of $J \;$ for stable orbits may be analysed. Although different
in details, the orbital motion with modified dispersion
relations is qualitatively similar to that in the standard case.

Consider a radially infalling motion. Then $J = 0 \;$. With no
loss of generality, we set $r = R \;$ and $u^r = 0 \;$, hence $X
= 0 \;$, initially at $t = \lambda = 0 \;$. This corresponds to
releasing a particle from rest at $r = R \;$. This also ensures
that ${\cal H}_0 < 0 \;$, hence that $G_{\mu \nu} u^\mu u^\nu <
0 \;$ and the trajectory is timelike initially. The particle
will fall radially towards the horizon at $r_h \;$ and, for the
function $f(X)$ given in equation (\ref{hhl}) with $z > 1 \;$,
it is clear from the general considerations given below equation
(\ref{umuunu}) that the initially timelike particle trajectory
will turn spacelike as $r \to r_{h +} \;$.

To see this explicitly, let $f = \alpha X^z \;$ with $\alpha > 0
\;$, and study the motion near $r_h \;$. The standard case may
be seen by setting $z = 1 \;$. The effects of modified
dispersion relations may be seen by setting $z > 1 \;$. 

Let $\rho = r - r_h \;$ and consider the limit $\rho \to 0_+ \;$
which corresponds to approaching the horizon. In this limit,
upto coefficients which are not important for present purposes,
we have $B \sim \rho \;$. Then $X^z \sim \rho^{- 1} \;$ and
\[
(u^r)^2 \; = \; B X f'^2 \; \sim \; \rho^{- \frac{z - 1}{z}}  
\; \; \; \; \; , \; \; \; \; \; \; 
(u^0)^2 \; \sim \; \rho^{- 2}  \; \; .
\]
Consider now the velocity, $v_{obs} \;$, of the particle as
measured by an observer who is at rest at $r = r_h + \rho
\;$. It is given by
\[
v_{obs} \; = \; c \; \sqrt{\frac{g_{rr}}{\vert g_{tt} \vert}} 
\; \frac{d r}{d t} \; = \; \frac{1}{B} \; \frac{u^r}{u^0} \; \;.
\]
See, for example, \cite{ll, ct} for more details on $v_{obs}
\;$. Thus, we have
\begin{equation}\label{vobs}
v^2_{obs} \; = \; \frac{(u^r)^2 }{p_0^2} 
\; \sim \; \rho^{- \frac{z - 1}{z}} 
\end{equation}
upto constant coefficients. Hence, in the limit $\rho \to 0_+
\;$, it follows that the velocity $v_{obs} \;$ of the particle,
as measured by an observer who is at rest at $r = r_h + \rho
\;$, approaches a constant if $z = 1 \;$ and diverges to
$\infty \;$ if $z > 1 \;$.

Another consequence of the modified dispersion relations may be
seen in the coordinate time required to reach the horizon, {\em
i.e.} in the value of $t$ as $\rho \to 0_+ \;$. The time $t$
required to reach $r = \rho \;$ is given, upto a finite additive
constant, by
\[
t = \int^\rho \frac{d r}{v^r} 
\]
where $v^r = \frac{d r}{d t} = \frac{u^r}{u^0} \;$. We have $v^r
\sim \rho^{\frac{z + 1}{2 z}} \;$ in the limit $\rho \to 0_+
\;$. From the above integral for $t \;$ we obtain, upto finite
additive constants, that 
\[
t \sim - \; ln \; \rho 
\; \; \; 
\; \; \; if  \; \; \; z = 1
\]
and 
\[
t \sim \rho^{\frac{z - 1}{2 z}} 
\; \; \; 
\; \; \; 
if  \; \; \; z > 1 \; \; . 
\]
It now follows that the horizon is reached only in the limit $t
\to \infty \;$ if $z = 1 \;$, which is well known, and that the
horizon is reached in a finite time $t \;$ if $z > 1 \;$ which
is a consequence of the modified dispersion relation in equation
(\ref{hhl}).

The above analysis is applicable also for the static spherically
symmetric solutions of Ho\v{r}ava's theory obtained in
\cite{blackholes}. The above qualitative features remain valid
in these cases also.

\vspace{2ex}

\centerline{ {\bf 5. Summary and conclusion} }

\vspace{2ex}

We first summarise the present work. We have studied the motion
of a particle obeying a modified dispersion relation. We used
super Hamiltonian formalism which is better suited for this
purpose. We considered the dispersion relation that incorporates
Ho\v{r}ava -- Lifshitz type anisotropic scaling symmetry,
characterised by a scaling exponent $z > 1 \;$. The standard
case corresponds to $z = 1 \;$.

We first studied the charged particle motion when a constant
electric field is present and find that, for $z > 1 \;$, the
particle speed diverges in the UV, namely, at high momentum.

We then studied the particle motion outside the horizon with
only gravitational fields present. For $z > 1 \;$, we find that
the particle speed, as measured by an observer locally at rest,
diverges as the particle approaches the horizon. Also, the
particle reaches the horizon in finite coordinate time $t \;$,
in contrast to the standard case where it requires infinite 
time. 

Our method is also applicable in the analysis of other types of
dispersion relations, postulated for various reasons
\cite{transplanck, dispersion}. A class of those studied in
\cite{transplanck}, in particular, are likely to exhibit the
same features as seen here.

We now conclude by mentioning several issues for further studis.
For the classical particles, we simply postulated the dispersion
relations and proposed a super Hamiltonian that incorporates it.
It is desireable to derive them formally. One may consider a
semiclassical probe wave packet, that may be expected to behave
like a classical particle. The dispersion relation for such a
wave packet may then be derived from a Lagrangian, of the type
recently constructed \cite{misc}, for its constituent fields
having anisotropic scaling symmetry.  In this context, see NOTE
added.

Another issue is the further evolution of the particle
approaching the horizon radially, or with small impact
parameter. Such a particle has superluminal speed as it
approaches the horizon. Does it enter the horizon? If it does,
can it then come out? It is not clear to us how best to address
this issue.

A related issue is whether a black hole forms in the near
head-on collison of two sufficiently high energy particles? In
the standard case, a black hole is expected to form once the two
particles are within the Schwarzschild radius of their center of
mass energy. Here, because of the modified dispersion relations,
their speeds are likely to be superluminal and it is not obvious
that a black hole will result.

Another related issue is whether a black hole forms when matter
collapses. In Ho\v{r}ava's theory, the equation of state for
matter will become $p = \frac{z}{d} \; \rho \;$ in the UV
\cite{cosmology}. There seems to be no obstacle for such matter
to collapse and form a black hole. However, see NOTE added.

We find that, as a particle falls towards the horizon, its speed
exceeds $c \;$ and grows without bound as it approaches the
horizon. It will be interesting to understand whether such a
particle can tell the presence of the horizon. This is not
possible in the standard case, but may be possible now since the
particle is not following the geodesics.

A more general question is whether the description of Physical
laws remain the same or not in the frame of a particle which
starts with subluminal speed and becomes superluminal due to its
modified dispersion relation. We find the question very
intersting, but it is not clear to us how to find the answer.

A recent paper \cite{hnew} has studied conformal structure and
possible Penrose type diagrams for spacetimes with anisotropic
scaling symmetry. It will be interesting to see if and how the
particle motions presented here may fit in, and/or may help
elucidate, such a spacetime structure.

\vspace{1ex}

{\bf NOTE added:} 

\vspace{1ex}

Two recent papers \cite{new1, new2} appeared while this paper
was being written. They start with a Lagrangian with anisotropic
scaling symmetry, and study the kinematics of the fields in the
geometric optics approximation using WKB methods. Among other
things, they also find a Hamiltonian and the corresponding
equations of motion, which agree with the present ones.

The paper \cite{new1} also studies collapse of matter. It
suggests that matter (dust, with pressure $p = 0 \;$) will first
collapse, but will bounce back near the singularity. Although it
is not proven yet, such a bounce may indicate that black holes
may not form in a collapse. However, in this paper, the same IR
equation of state ($p = 0 \;$) is used even in the UV near the
singularity and it was found that matter has no effect on the
bounce. But in Ho\v{r}ava's theory, the equation of state of
matter changes in the UV, generically to $p = \frac{z}{d} \;
\rho \;$. Such an equation of state will then have non trivial
effect on the bounce, see S. K. Rama in \cite{cosmology}. So it
is not clear to us if the bounce in \cite{new1} will still be
present if the correct equation of state in the UV is taken into
account.


\vspace{3ex}

{\bf Acknowledgement: } We thank G. Date for a discussion.



\begin{thebibliography}{999}


\bibitem{h1}
P.~Horava,
Phys.\ Rev.\  D {\bf 79}, 084008 (2009) 
[arXiv:0901.3775 [hep-th]].
See also 
P.~Horava,
JHEP {\bf 03}, 020 (2009)
[arXiv:0812.4287 [hep-th]]; 
P.~Horava,
Phys.\ Rev.\  Lett. {\bf 102}, 161301 (2009)
[arXiv:0902.3657 [hep-th]].


\bibitem{cosmology}
T.~Takahashi and J.~Soda,
Phys.\ Rev.\ Lett.\  {\bf 102}, 231301 (2009)
[arXiv:0904.0554 [hep-th]].
G.~Calcagni,
JHEP {\bf 09}, 112 (2009)
[arXiv:0904.0829 [hep-th]].
E.~Kiritsis and G.~Kofinas,
Nucl.\ Phys.\  B {\bf 821}, 467 (2009)
[arXiv:0904.1334 [hep-th]].
S.~Mukohyama,
JCAP {\bf 06}, 001 (2009)
[arXiv:0904.2190 [hep-th]].
R.~Brandenberger,
Phys.\ Rev.\  D {\bf 80}, 043516 (2009)
[arXiv:0904.2835 [hep-th]].
Y.~S.~Piao,
arXiv:0904.4117 [hep-th].
X.~Gao,
arXiv:0904.4187 [hep-th].
S.~Mukohyama, K.~Nakayama, F.~Takahashi and S.~Yokoyama,
Phys.\ Lett.\  B {\bf 679}, 6 (2009)
[arXiv:0905.0055 [hep-th]].
R.~G.~Cai, B.~Hu and H.~B.~Zhang,
Phys.\ Rev.\  D {\bf 80}, 041501 (2009)
[arXiv:0905.0255 [hep-th]].
S.~Kalyana Rama,
Phys.\ Rev.\  D {\bf 79}, 124031 (2009)
[arXiv:0905.0700 [hep-th]].
B.~Chen, S.~Pi and J.~Z.~Tang,
arXiv:0905.2300 [hep-th].
M.~Minamitsuji,
arXiv:0905.3892 [astro-ph.CO].
S.~Mukohyama,
Phys.\ Rev.\  D {\bf 80}, 064005 (2009)
[arXiv:0905.3563 [hep-th]].
A.~Wang and Y.~Wu,
JCAP {\bf 07}, 012 (2009)
[arXiv:0905.4117 [hep-th]].
S.~Nojiri and S.~D.~Odintsov,
arXiv:0905.4213 [hep-th].
A.~Wang and R.~Maartens,
arXiv:0907.1748 [hep-th].
T.~Kobayashi, Y.~Urakawa and M.~Yamaguchi,
arXiv:0908.1005 [astro-ph.CO].
S.~Maeda, S.~Mukohyama and T.~Shiromizu,
arXiv:0909.2149 [astro-ph.CO].
C.~G.~Boehmer and F.~S.~N.~Lobo,
arXiv:0909.3986 [gr-qc].


\bibitem{blackholes}
H.~Lu, J.~Mei and C.~N.~Pope,
Phys.\ Rev.\ Lett.\  {\bf 103}, 091301 (2009)
[arXiv:0904.1595 [hep-th]].
H.~Nastase,
arXiv:0904.3604 [hep-th].
R.~G.~Cai, L.~M.~Cao and N.~Ohta,
Phys.\ Rev.\  D {\bf 80}, 024003 (2009)
[arXiv:0904.3670 [hep-th]].
R.~G.~Cai, Y.~Liu and Y.~W.~Sun,
JHEP {\bf 06}, 010 (2009)
[arXiv:0904.4104 [hep-th]].
E.~O.~Colgain and H.~Yavartanoo,
JHEP {\bf 08}, 021 (2009)
[arXiv:0904.4357 [hep-th]].
S.~S.~Kim, T.~Kim and Y.~Kim,
arXiv:0907.3093 [hep-th].
J.~Z.~Tang and B.~Chen,
arXiv:0909.4127 [hep-th].


\bibitem{motion}
R.~A.~Konoplya,
Phys.\ Lett.\  B {\bf 679}, 499 (2009)
[arXiv:0905.1523 [hep-th]].
J.~Chen and Y.~Wang,
arXiv:0905.2786 [gr-qc].
S.~Mukohyama,
JCAP {\bf 09}, 005 (2009)
[arXiv:0906.5069 [hep-th]].
T.~Harko, Z.~Kovacs and F.~S.~N.~Lobo,
Phys.\ Rev.\  D {\bf 80}, 044021 (2009)
[arXiv:0907.1449 [gr-qc]].
T.~Harko, Z.~Kovacs and F.~S.~N.~Lobo,
arXiv:0908.2874 [gr-qc].


\bibitem{misc}
M.~Visser,
Phys.\ Rev.\  D {\bf 80}, 025011 (2009)
[arXiv:0902.0590 [hep-th]].
T.~P.~Sotiriou, M.~Visser and S.~Weinfurtner,
Phys.\ Rev.\ Lett.\  {\bf 102}, 251601 (2009)
[arXiv:0904.4464 [hep-th]].
B.~Chen and Q.~G.~Huang,
arXiv:0904.4565 [hep-th].
T.~Nishioka,
arXiv:0905.0473 [hep-th].
J.~Kluson,
Phys.\ Rev.\  D {\bf 80}, 046004 (2009)
[arXiv:0905.1483 [hep-th]].
T.~P.~Sotiriou, M.~Visser and S.~Weinfurtner,
arXiv:0905.2798 [hep-th].
G.~Calcagni,
arXiv:0905.3740 [hep-th].
M.~Sakamoto,
Phys.\ Rev.\  D {\bf 79}, 124038 (2009)
[arXiv:0905.4326 [hep-th]].
R.~Iengo, J.~G.~Russo and M.~Serone,
arXiv:0906.3477 [hep-th].

\bibitem{romero}
J.~M.~Romero, V.~Cuesta, J.~A.~Garcia and J.~D.~Vergara,
{\em Conformal Anisotropic Mechanics}, 
arXiv:0909.3540 [hep-th].

\bibitem{mtw}
C. W. Misner, K. S. Thorne, and J. A. Wheeler, 
{\em Gravitation}, W. H. Freeman and Company (1973); 
see sections (21.1) and (33.5), and exercises (25.2) and
(25.11).

\bibitem{transplanck}
W.~G.~Unruh,
Phys.\ Rev.\  D {\bf 51}, 2827 (1995).
S.~Corley and T.~Jacobson,
Phys.\ Rev.\  D {\bf 54}, 1568 (1996)
[arXiv:hep-th/9601073].
J.~Martin and R.~H.~Brandenberger,
Phys.\ Rev.\  D {\bf 63}, 123501 (2001)
[arXiv:hep-th/0005209].
R.~H.~Brandenberger and J.~Martin,
Mod.\ Phys.\ Lett.\  A {\bf 16}, 999 (2001)
[arXiv:astro-ph/0005432].
J.~C.~Niemeyer and R.~Parentani,
Phys.\ Rev.\  D {\bf 64}, 101301 (2001)
[arXiv:astro-ph/0101451].

\bibitem{dispersion}
G.~Amelino-Camelia,
Int.\ J.\ Mod.\ Phys.\  D {\bf 11}, 35 (2002)
[arXiv:gr-qc/0012051].
N.~R.~Bruno, G.~Amelino-Camelia and J.~Kowalski-Glikman,
Phys.\ Lett.\  B {\bf 522}, 133 (2001)
[arXiv:hep-th/0107039].
J.~Magueijo and L.~Smolin,
Phys.\ Rev.\ Lett.\  {\bf 88}, 190403 (2002)
[arXiv:hep-th/0112090].
J.~Kowalski-Glikman and S.~Nowak,
Phys.\ Lett.\  B {\bf 539}, 126 (2002)
[arXiv:hep-th/0203040].
J.~Lukierski and A.~Nowicki,
Int.\ J.\ Mod.\ Phys.\  A {\bf 18}, 7 (2003)
[arXiv:hep-th/0203065].
J.~Kowalski-Glikman and S.~Nowak,
Int.\ J.\ Mod.\ Phys.\  D {\bf 12}, 299 (2003)
[arXiv:hep-th/0204245].
S.~Kalyana Rama,
Mod.\ Phys.\ Lett.\  A {\bf 18}, 527 (2003)
[arXiv:hep-th/0209129].

\bibitem{ll}
L. D. Landau and E. M. Lifshitz, 
{\em The Classical Theory of Fields}, 
Pergamon press (1975), IV edition;
see sections (84) and (88). 

\bibitem{ct}
P.~Crawford and I.~Tereno,
Gen.\ Rel.\ Grav.\  {\bf 34}, 2075 (2002)
[arXiv:gr-qc/0111073].

\bibitem{hnew}
P.~Horava and C.~M.~Melby-Thompson,
{\em Anisotropic Conformal Infinity},
arXiv:0909.3841 [hep-th].

\bibitem{new1}
T.~Suyama,
{\em Notes on Matter in Horava-Lifshitz Gravity}, 
arXiv:0909.4833 [hep-th].

\bibitem{new2}
D.~Capasso and A.~P.~Polychronakos,
{\em Particle Kinematics in Horava-Lifshitz Gravity},
arXiv:0909.5405 [hep-th].

\end{thebibliography}
\end{document}